\documentclass[12pt]{article}
\usepackage[dvips]{color}
\usepackage{amsmath}
\usepackage{graphicx}
\usepackage{subfigure}
\usepackage{epsfig}
\usepackage{amsmath}
\usepackage{cite}
\usepackage{color}
\usepackage{subfigure}

\begin{document}
\begin{center}
\Large{\bf Reissner-Nordstr$\ddot{o}$m black holes surrounded by perfect fluid dark matter: testing the viability of weak gravity conjecture and weak cosmic censorship conjecture simultaneously}\\
\small \vspace{1cm} {\bf Jafar Sadeghi$^{a}$\footnote {Email:~~~pouriya@ipm.ir}}, \quad
{\bf Saeed Noori Gashti$^{a}$\footnote {Email:~~~saeed.noorigashti@stu.umz.ac.ir}} \quad
 \\
\vspace{0.5cm}$^{a}${Department of Physics, Faculty of Basic
Sciences,
University of Mazandaran\\
P. O. Box 47416-95447, Babolsar, Iran}\\
\small \vspace{1cm}
\end{center}
\begin{abstract}
A possible violation of the weak gravity conjecture (WGC) by cosmic censorship is one of the major challenges in the field of general relativity. The cosmic censorship is a fundamental principle that ensures the consistency of the theory of gravity. According to this principle, a charged black hole in four dimensions cannot violate the weak cosmic censorship conjecture (WCCC) under normal conditions. However, in this paper, we explore the possibility of reconciling the WGC and the WCCC by considering Reissner-Nordstr$\ddot{o}$m (R-N) black holes embedded in perfect fluid dark matter (PFDM) in asymptotically flat spacetimes. These two conjectures are seemingly unrelated, but a recent proposal suggested that they are connected surprisingly. In particular, We argue a promising class of valid counterexamples to the WCCC in the four-dimensional Einstein-Maxwell theory, considering a charged black hole when  WGC is present. We demonstrate that by imposing certain constraints on the parameters of the metric, the WGC and the WCCC can be compatible. Furthermore, we investigate the properties of the charged black hole in the presence of PFDM for $Q > M$ and present some intriguing figures to test the validity of the WGC and the WCCC simultaneously. When PFDM is absent ($\gamma=0$), the RN black hole either has two event horizons if $Q^2/M^2\leq 1$ or none if $Q^2/M^2> 1$. The second scenario results in a naked singularity, which contradicts the WCCC. But when PFDM is present ($\gamma\neq 0$), the RN black hole has event horizons with regard to Q and M. This implies that the singularity is always covered, and the WGC and the WCCC are fulfilled. Furthermore, we demonstrate that there is a critical value of $\gamma$, called $\gamma_{ext}$, that makes the RN black hole extremal when $\gamma=\gamma_{ext}$. In this situation, the black hole has an event horizon, and the WGC and the WCCC are still fulfilled. We infer that PFDM can make the WGC and the WCCC compatible with the RN black hole and that the WGC and the WCCC agree with each other when PFDM is present.
\end{abstract}
\tableofcontents
\section{Introduction}
Quantum gravity is a fascinating topic that has attracted the attention of many researchers from different perspectives. The swampland program is a research effort to find the universal principles that any consistent theory of quantum gravity must obey. It is based on the idea that not all low-energy effective field theories can be embedded in a quantum theory of gravity, such as string theory. The swampland program tries to identify the criteria that distinguish the theories that can be part of the string landscape from those that belong to the swampland. The motivation for the swampland program comes from various sources, such as black hole physics, AdS/CFT correspondence, and string theory constructions. The swampland program aims to shed light on the nature of quantum gravity and its implications for cosmology and particle physics. The AdS/CFT correspondence is a duality that relates a gravitational theory in anti-de Sitter (AdS) space to a conformal field theory (CFT) on its boundary. It is a powerful tool to study quantum gravity and its holographic nature. The swampland program is an attempt to find the universal constraints that any consistent theory of quantum gravity must satisfy. One of the sources of motivation for the swampland program is the AdS/CFT correspondence, as it provides concrete examples of quantum gravity theories and their low-energy limits. The AdS/CFT correspondence can also be used to test some of the swampland conjectures, such as the bound on the number of massless modes, the weak gravity conjecture, and the distance conjecture\cite{1,2,3,4,5,6}.\\\\ A key criterion in the swampland program is the absence of global symmetries in quantum gravity, while gauge symmetries are allowed \cite{1,2,3,4,5,6}. This criterion leads to the weak gravity conjecture (WGC), which states that there must exist particles whose charge-to-mass ratio is greater than one, i.e. $q/m > 1$, in any quantum theory of gravity \cite{4,5,6}. In this way, gravity is the weakest force among all the interactions. The WGC is one of the several conjectures in the swampland program that help to identify effective field theories that are consistent with quantum gravity. For more details, see\cite{1,2,3,4,5,6,7,8}. Also for further study about the swampland program in various cosmological concepts such as the physics of black holes. thermodynamics, black brane, cosmological inflation, etc, you can see\cite{a,b,c,d,e,f,g,h,i,j,k,l,m,n,o,p,q,r,s,t,u,v,w,x,y,z,aa,bb,cc,dd,ee,ff,gg,hh,ii,jj,kk,ll,mm,nn,oo,pp,qq,rr,ss,tt,uu,vv,ww,xx,yy,zz,aaa,bbb,ccc,ddd,eee,fff,ggg,hhh,iii}. Another important concept in theoretical physics is the weak cosmic censorship conjecture (WCCC), which was proposed by Penrose to avoid the paradoxes caused by the existence of singularities in general relativity\cite{9}. The WCCC asserts that the singularities formed by gravitational collapse are always hidden behind event horizons, thus preserving the causal structure and predictability of the theory. However, the WGC and the WCCC seem to be incompatible with each other in the case of the Reissner–Nordström (RN) black hole, which is a solution of the Einstein–Maxwell equations describing a charged black hole. When $Q > M$, where Q is the charge and M is the mass of the black hole, the RN solution violates the WCCC, as the singularity becomes naked and visible to distant observers. On the other hand, when the RN black hole decays to the extremal state, where Q = M, the energy conservation implies that there are decay products whose charge-to-mass ratio is greater than one. In this case, the WGC is satisfied, but the WCCC is violated, as these products cannot be black holes, but rather particles\cite{10}. This is one of the main challenges for the WGC.\\\\
Recently, researchers have studied WCCC for the RN-dS black hole with PFDM. They demonstrated that the presence of dark matter and a cosmological constant prevents the overcharging of this black hole, a scenario not possible in a vacuum. The only condition under which overcharging could occur is if there is an exact equilibrium between the influences of dark matter and the cosmological constant. Moreover, there is a specific limit where the influence of the cosmological constant becomes dominant, ensuring the black hole cannot be overcharged and thus, the WCCC is maintained. This conclusion is consistent across both linear and non-linear accretion scenarios, indicating that these cosmic factors act as safeguards against the overcharging of black holes\cite{10000}.
In this paper, we address this challenge by considering the effects of perfect fluid dark matter (PFDM) on the charged black holes. PFDM is a model of dark matter that behaves as a perfect fluid with a constant equation of state\cite{13,11,12,14,15,16,17}. We study the properties of the RN black holes surrounded by PFDM in asymptotically flat spacetimes. We show that by imposing certain constraints on the parameters of the metric, such as the mass, the charge, and the PFDM density, we can achieve compatibility between the Mild WGC and the WCCC. We also investigate the case of $Q > M$ and present some interesting points to test the validity of the Mild WGC and the WCCC simultaneously.
Our paper is motivated by the recent observational evidence of the accelerated expansion of the universe\cite{11,12}, which suggests the existence of a repulsive gravitational force due to a negative pressure on cosmological scales. One possible explanation for this phenomenon is the presence of a cosmological constant, which corresponds to the vacuum energy in Einstein's equations. However, the cosmological constant also affects the properties of the black holes and their compatibility with the Mild WGC and the WCCC. Therefore, we explore how the charged black holes with the PFDM model can account for the consistency of the quantum gravity due to consistent with Mild WGC and its effect on the validation of the Mild WGC and WCCC. On the other hand, The motivation of the article is to explore the possibility of resolving a theoretical conflict between two conjectures (Mild WGC and the WCCC simultaneously), with respect to the effects of PFDM  that has important implications for quantum gravity and black hole physics. This is a challenging and interesting research topic that could shed light on some fundamental aspects of nature.\\\\
The paper is organized as follows. In section 2, we review the basic features of the RN black hole surrounded by PFDM and the effects of the Mild WGC and the WCCC. In section 3, we study the consistency between Mild WGC and WCCC for the mentioned model viz the RN black hole surrounded by PFDM. In section 4, we analyze the properties of the parameters and the constraints on them and present some interesting figures to test the validity of the Mild WGC and the WCCC simultaneously. In section 5, we summarize our results and conclude with some remarks.
\section{The Model}
Reissner-Nordström black holes surrounded by perfect fluid dark matter are electrically charged black holes that are influenced by the presence of a hypothetical form of matter that accounts for most of the mass in the universe.
They have interesting thermodynamic and phase transition properties that can be studied by extending the system's phase space.
Some researchers have also tested the weak cosmic censorship conjecture for these black holes, which states that the singularity inside a black hole cannot be exposed to an external observer. The following action describes the gravitational theory that is minimally coupled with a gauge field in the presence of perfect fluid dark matter (PFDM)\cite{133,144,155,166},
\begin{equation}\label{eq1}
S=\int d^4x\sqrt{-g}\bigg[\frac{1}{16\pi G}R+\frac{1}{4}F^{\mu\nu}F_{\mu\nu}+\mathcal{L}_{DM}\bigg],
\end{equation}
where $g$ as the determinant of the metric tensor $[g=det_(g_{ab})]$, $R$ as the scalar curvature, $G$ as the gravitational constant, and $F_{\mu\nu}$ as the electromagnetic tensor derived from the gauge potential $A_{\mu}$. $L_{DM}$ is the Lagrangian density that describes the PFDM. By applying the principle of least action, one can derive the Einstein field equations as follows,
\begin{equation}\label{eq2}
\begin{split}
R_{\mu\nu}-\frac{1}{2}g_{\mu\nu}R= 8\pi G(T^{M}_{\mu\nu}-T^{DM}_{\mu\nu})\equiv -8\pi GT_{\mu\nu},
\end{split}
\end{equation}
\begin{equation}\label{eq3}
\begin{split}
&F^{\mu\nu}_{\nu}=0,\\
&F^{\mu\nu;\alpha}+F^{\nu\alpha;\mu}+F^{\alpha\mu;\nu}=0.
\end{split}
\end{equation}
We denote $T^M_{\mu\nu}$ as the energy-momentum tensor of the ordinary matter and $T^{DM}_{\mu\nu}$ as the energy-momentum tensor of the PFDM, as defined in\cite{133,144,155,166}. So,
\begin{equation}\label{eq4}
\begin{split}
&T^{\mu}_{\nu}=g^{\mu\nu}T_{\mu\nu},\\
&T^{t}_{t}=-\rho,\hspace{0.5cm}T^{r}_{r}=T^{\theta}_{\theta}=T^{\phi}_{\phi}=P.
\end{split}
\end{equation}
where one can assume that $T^{r}_{r}=T^{\theta}_{\theta}=T^{\phi}_{\phi}= T^{t}_{t} (1-\delta)$, where $\delta$ is a constant, based on\cite{133,144,155,166}. The expression for the Reissner-Nordström metric in the context of PFDM is given below,
\begin{equation}\label{eq5}
\begin{split}
ds^{2}=-f(r)dt^{2}+f(r)^{-1}dr^{2}+r^{2}(d\theta^{2}+\sin^2\theta d\phi^{2}),
\end{split}
\end{equation}
where,
\begin{equation}\label{eq6}
f(r)=1-\frac{2M}{r}+\frac{Q^2}{r^{2}}+\frac{\gamma}{r}\ln(\frac{r}{\gamma}).
\end{equation}
The parameters M, Q, and $\gamma$ represent the mass, charge, and PFDM contribution of the black hole, respectively. If the PFDM is absent ($\gamma$ = 0), the above space-time metric reduces to the Reissner-Nordström black hole.
The values of $\gamma$ can be both positive and negative and here, we consider both of them in this work.
\section{Mild WGC $\&$ WCCC}
We consider the metric of a R-N black hole surrounded by PFDM in asymptotic flatness. By solving $f(r) = 0$, and using the substitution $x = 1/r$, we can find the location of the event horizon for the black hole. Generally, we can find the location of the event horizons for a charged black hole by solving f(r)=0 and get $r_{\pm} =M \pm \sqrt{M^2-Q^2}$, where M is the mass and Q is the charge of the black hole. If $Q>M$, then the black hole does not have any event horizon, and its singularity is exposed to the outside observers. This is called a naked singularity, and it violates the weak cosmic censorship conjecture (WCCC), which states that singularities should be hidden behind event horizons. So, we rewrite the equation (6) as follows,
\begin{equation}\label{eq7}
1-2 M x+Q^2 x^2=-\gamma x \ln (\frac{1}{x \gamma})
\end{equation}
Eq. (7) is a difficult equation to solve. We use some plots and analyze various scenarios to find their solutions. We start with the case of asymptotically flat space, where $\gamma<0$ or $\gamma>0$. Based on Eq. (7), we sketch two curves $\Upsilon_L=1-2 M x+Q^2 x^2$ and $\Upsilon_R = -\gamma x \ln (\frac{1}{x \gamma})$ and label them as Fig. 1. We consider two situations: $Q^2/M^2\leq1$ and $Q^2/M^2>1$. In Fig. 1c, when $\Upsilon_R = 0$, the curve $\Upsilon_L$ crosses the $x$ axis at two points $(x_+,x_-)$. These are the inner and outer event horizons of the standard RN black hole. We use Eq. (7) and the condition that the curves of $\Upsilon_R$ and $\Upsilon_L$ are tangent to each other at the point $x_0$ to find the values of $x_0$ and $\gamma_{ext}$. This means that the slopes of the two curves are equal at $x_0$, and we can use this to solve for $\gamma_{ext}$. Then, we can calculate the value of $\gamma_{ext}$ and $x_0$ with the following expressions,
\begin{equation}\label{eq8}
1-2 M x_0+Q^2 x_0^2=-\gamma x_0 \ln (\frac{1}{x_0 \gamma}),
\end{equation}
and
\begin{equation}\label{eq9}
-\gamma-2 M+2 Q^2 x_0=-\gamma \ln (\frac{1}{x_0 \gamma}).
\end{equation}
Using the equations (8) and (9), we will have,
\begin{equation}\label{eq10}
x_0=\frac{\gamma\pm \sqrt{\gamma^2+4Q^2}}{2Q^2}
\end{equation}
Also, we have the $\gamma_{ext}$ by using the expansion of the logarithm up to two terms and a series of simplifications,
\begin{equation}\label{eq10}
\gamma_{ext}=\frac{\bigg[2 Q^2 x_0^3\pm 2 \sqrt{x_0^2 \bigg(\big(-2 M x_0+Q^2 x_0^2+3\big){}^2-3\bigg)}\bigg]+2 x_0 \big(3-2 M x_0\big)}{6 x_0^2}
\end{equation}
We can obtain the extremal state of the black hole by keeping (Q, M, $\gamma$) constant and increasing $Q^2/M^2$ until it reaches the value of $(Q^2/M^2)_{ext}$. This is the condition for the black hole to have a single event horizon. We can also use some approximation for the RN extremal black hole with PFDM, by fixing $\gamma$, Q, M and using Eq. (6) and (7). This gives us the relation for the extremal black hole with PFDM, which is,
\begin{equation}\label{eq11}
(\frac{M^2}{Q^2})_{ext}=\frac{Q^2 \bigg(\frac{6 \gamma \big(\gamma\pm \sqrt{\gamma^2+4 Q^2}\big)}{Q^2}+\frac{\gamma \big(\gamma\pm \sqrt{\gamma^2+4 Q^2}\big)^3}{4 Q^4}-\frac{3 \gamma^2 \big(\gamma\pm \sqrt{\gamma^2+4 Q^2}\big)^2}{4 Q^4}-1\bigg)}{\gamma \bigg(\gamma\pm \sqrt{\gamma^2+4 Q^2}\bigg)^2}
\end{equation}
With a little simplification and a series of straightforward calculations, the relation $(\frac{Q^2}{M^2})_{ext}$ is rewritten in the following form,
\begin{equation}\label{eq12}
(\frac{Q^2}{M^2})_{ext_{1,2}}=\pm 4\gamma(1\mp\frac{\gamma^2}{Q^2}+...)
\end{equation}
So, finally for the second term (the negative part behind the parentheses) with $\gamma\leq -1/4$ and $Q>0$, we will have,
\begin{equation}\label{eq13}
(\frac{Q^2}{M^2})_{ext}=1+\varepsilon,
\end{equation}
where $\varepsilon$ is a constant positive value. From the above concepts, we want to investigate whether the weak gravity conjecture (WGC) and the weak cosmic censorship conjecture (WCCC) are satisfied for the Reissner-Nordström (RN) black hole with (PFDM) in both normal and extremal states.
\section{Discussion and result}
We examine the graph of $\Upsilon_L$ and find that it has a minimum value of $(Q^2-M^2)/Q^2$ at the point $x_{min} = M/Q^2$. We also look at Fig. 1c, where we plot $\Upsilon_R$ and $\Upsilon_L$ for the case of $Q^2/M^2 \leq 1$. We observe that the two curves intersect at two points $(x_1, x_2)$. This implies that the RN black hole with PFDM has an outer event horizon $r_1 = 1/x_1$ that is larger than the outer event horizon of the ordinary RN black hole $r_+ = 1/x_+$, and an inner event horizon $r_2 = 1/x_2$ that is smaller than the inner event horizon of the ordinary RN black hole $r_- = 1/x_-$. On the other hand, in Fig. 1a, we plot $\Upsilon_R$ and $\Upsilon_L$ for the case of $Q^2/M^2 > 1$. We notice that the curve $\Upsilon_L$ does not cross the $x$-axis, which means that the ordinary RN black hole has no event horizon and exposes its singularity to the outside. This is called a naked singularity, and it violates the WCCC and the Mild WGC is not valid. However, in the presence of PFDM, the curves $\Upsilon_R$ and $\Upsilon_L$ still intersect at two points $x_1$ and $x_2$, even when $Q^2/M^2 > 1$. This means that the RN black hole with PFDM has two event horizons that hide its singularity from the observers. Therefore, the WCCC and the Mild WGC are not violated, and they are consistent with each other. In Fig. 1b, the curves $\Upsilon_R$ and $\Upsilon_L$ touch each other at a single point. In this case, we can define the extremal state of the black hole at the point $x_0$, where $Q^2/M^2 > 1$ and $\gamma = \gamma_{ext}$. When the parameters M and Q are fixed, if $\gamma > \gamma_{ext}$, then the black hole becomes a naked singularity and violates the WCCC. If $\gamma < \gamma_{ext}$, then the black hole has two event horizons and satisfies the WCCC. If $\gamma = \gamma_{ext}$, then the black hole has a single event horizon $x_0=1$ and is in the extremal state. This situation is clearly shown in Fig. 2. We plot $\Upsilon_L=1-2 M x+Q^2 x^2$ and $\Upsilon_R = -\gamma x \ln (\frac{1}{x \gamma})$ in Fig. 1, using Eq. (7). We consider two cases: $Q^2/M^2\leq1$ and $Q^2/M^2> 1$. In Fig. 1c, when $\Upsilon_R = 0$, the curve $\Upsilon_L$ crosses the $x$-axis at two points $(x_+, x_-)$. These are the inner and outer event horizons of the standard RN black hole. We investigate the conditions for the compatibility of the Mild WGC and the weak cosmic censorship conjecture (WCCC) for the Reissner-Nordström (RN) black hole with (PFDM). We use Yes to indicate that the two conjectures are consistent with each other, and No to indicate that they are not. We summarize the results in Table 1. In summary, in studying R-N black holes, it's noted that without PFDM, black holes with charge $(Q)$ greater than mass $(M)$, $(Q > M)$, lack the event horizon, revealing singularities. However, with PFDM, these black holes exhibit singularity and event horizon. This condition supports the simultaneous compatibility of the WGC and the WCCC, which is not possible in typical conditions. Also, at the extremality limit, both conditions are met. For WCCC to hold, the PFDM parameter must be within a certain range and also it must have a special range for WGC. The intersection of these ranges indicates the conditions under which both conjectures are satisfied for a charged black hole. Our research found a common range for the PFDM parameter that ensures the coexistence of WGC and WCCC, along with the presence of event horizon, satisfying the condition $(Q>M)$. The compatibility is shown using parameters $(Y_L)$ and $(Y_R)$, with constraints like those in Eq. (14). For instance, WGC is met for $(\gamma\leq -0.25)$ with respect to $Q>M$ also it is compatible with WCCC. Conversely, for $(\gamma>-0.25)$, WCCC holds, but not WGC (Eq. (14)), as seen in the intersection of $(Y_L)$ and $(Y_R)$. We employed plots to visually depict the parameter space where WGC and WCCC conditions align. These plots clarify the regions of compatibility, offering a graphical representation of our theoretical results. Furthermore, we utilized the Mild WGC, which posits gravity as the weakest force, allowing for a wider range of scenarios than the Strong WGC, which is more restrictive and particle-specific. Our studies reveal that under certain conditions, such as the inclusion of PFDM, WGC, and WCCC can coexist without contradiction, ensuring consistent physical predictions for black holes and their thermodynamics.

\newpage
\begin{figure}[h!]
 \begin{center}
 \subfigure[]{
 \includegraphics[height=5.5cm,width=5.5cm]{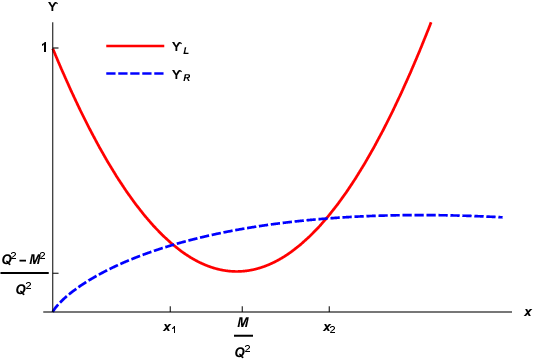}
 \label{1a}}
 \subfigure[]{
 \includegraphics[height=5.5cm,width=5.5cm]{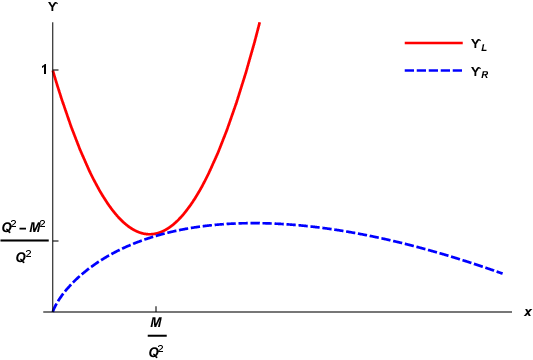}
 \label{1b}}
 \subfigure[]{
 \includegraphics[height=5.5cm,width=5.5cm]{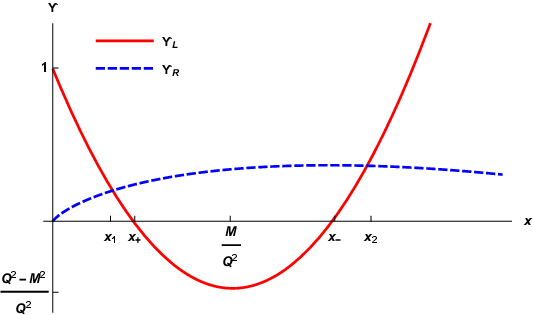}
 \label{1c}}
  \caption{\small{The plot of $\Upsilon-x$. $\Upsilon_L=1-2 M x+Q^2 x^2$ and $\Upsilon_R = -\gamma x \ln (\frac{1}{x \gamma})$ with respect to $\gamma<0$ }}
 \label{1}
 \end{center}
 \end{figure}

\begin{figure}[h!]
 \begin{center}
 \subfigure[]{
 \includegraphics[height=7.5cm,width=10.5cm]{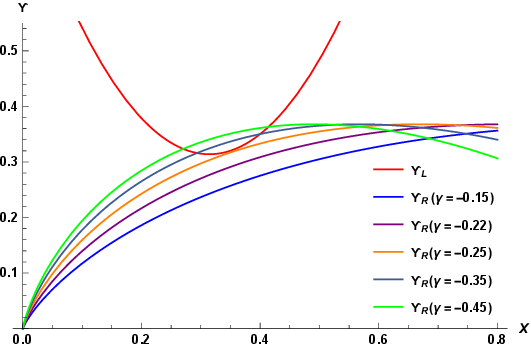}
 \label{2a}}
  \caption{\small{The plot of $\Upsilon-x$. $\Upsilon_L=1-2 M x+Q^2 x^2$ and $\Upsilon_R = -\gamma x \ln (\frac{1}{x \gamma})$ with respect to $Q>M$ and $\gamma<0$ }}
 \label{2}
 \end{center}
 \end{figure}

\section{Conclusions}
In this paper, the compatibility of the Mild weak gravity conjecture (WGC) and the weak cosmic censorship conjecture (WCCC) was explored by considering Reissner-Nordstr$\ddot{o}$m (R-N) black holes surrounded by perfect fluid dark matter (PFDM) in asymptotically flat spacetimes. The Mild WGC is a conjecture that imposes a lower bound on the charge-to-mass ratio in a quantum theory of gravity. The WCCC is a conjecture that states that singularities formed by gravitational collapse are hidden behind event horizons, preserving the determinism and predictability of general relativity. These two conjectures are seemingly unrelated, but a recent proposal suggested that they are connected surprisingly. In particular, it was argued a promising class of valid counterexamples to the WCCC in the four-dimensional Einstein-Maxwell theory, considering charged black holes when Mild WGC was present. This implies that the MILD WGC can protect the WCCC from violations, and that cosmic censorship is a robust principle that can withstand the challenges posed by the Mild WGC.
In this paper, we extended this analysis to the case of R-N black holes embedded in PFDM in asymptotically flat spacetimes. PFDM is a model of dark matter that behaves as a perfect fluid with a constant equation of state. We considered the effects of PFDM on the properties of the charged black hole, such as the horizon radius with respect to Mild WGC and WCCC. We imposed certain constraints on the parameters, such as the mass, the charge, and the PFDM to demonstrate that by choosing appropriate values of these parameters, the Mild WGC and the WCCC can be compatible. Moreover, we investigated the properties of the charged black hole in the presence of PFDM for $Q > M$, where Q is the charge and M is the mass of the black hole. We presented some interesting figures to test the validity of the Mild WGC and the WCCC simultaneously. We showed that the Mild WGC and WCCC can be satisfied for a special range of values of Q, M, and PFDM. The main goal of this study is to examine the effects of PFDM on the Reissner-Nordström (RN) black hole and its implications for the weak gravity conjecture (WGC) and the weak cosmic censorship conjecture (WCCC). We use a graphical method to plot the functions $\Upsilon_L$ and $\Upsilon_R$, which represent the left-hand side and the right-hand side of the equation of motion for the RN black hole with PFDM, respectively. We consider two cases: $Q^2/M^2\leq 1$ and $Q^2/M^2> 1$, where Q and M are the charge and mass of the black hole, respectively. We also vary the parameter $\gamma$, which affects of the PFDM on Mild WGC and WCCC.
We find that, in the absence of PFDM ($\gamma=0$), the RN black hole has two event horizons for $Q^2/M^2\leq 1$, and no event horizon for $Q^2/M^2> 1$. The latter case leads to a naked singularity, which violates the WCCC. However, in the presence of PFDM ($\gamma\neq 0$), the RN black hole has event horizons, with regard to the values of Q and M. This means that the singularity is always hidden, and the Mild WGC and the WCCC are satisfied. Moreover, we show that there exists a critical value of $\gamma$, denoted by $\gamma_{ext}$, such that the RN black hole becomes extremal when $\gamma=\gamma_{ext}$. In this case, the black hole has an event horizon, and the Mild WGC and the WCCC are still satisfied. We conclude that PFDM can restore the compatibility of the Mild WGC and the WCCC for the RN black hole and that the Mild WGC and the WCCC are consistent with each other in the presence of PFDM. We also present a table that summarizes results for different values $\gamma$.
The results of this paper have important implications for the field of general relativity and the quest for a quantum theory of gravity. They also suggest that cosmic censorship is a robust principle that can withstand the challenges.\\
The Weak Gravity Conjecture (WGC) is a new idea in theoretical physics. While we have not yet obtained significant observational evidence for the WGC, recently it plays a crucial role in understanding cosmic structure and particle physics. This includes phenomena such as physics of black hole, thermodynamics, dark energy, black brane and so on. Despite the scarcity of directly observable data related to these concepts, the WGC has already made some predictions. We anticipate that research in this area will accelerate in the coming years, filling gaps and providing a deeper understanding of fundamental physics. Of course, there are observational signatures for each of the examined concepts, which we briefly mention below. We hope that more observables related to combining concepts and proving the correctness of the issue will be identified in the near future.\\ We face with some concepts related to WGC such as:
1. Neutrinos: The WGC implies that Standard Model neutrinos must be electrically neutral.
2. Quantization of electric charge: The electric charge in the Standard Model must be quantized.
3. Masslessness of the photon: The photon must remain massless\cite{10,10'}.\\
To test the WCCC, scientists explore the following areas:
1. Black hole shadows: Investigating shadows cast by black holes, especially those with some properties (e.g., charged black holes or modified gravity theories).
2. Photon rings and lensing rings: These features relate to dark matter presence and black hole spacetime geometry.\\ Additionally, researchers study the Reissner-Nordstr$\ddot{o}$m, Kerr–Newman black holes and so on:
1. Shadow size: Increases with black hole charge.
2. Specific intensities: Charge affects observed intensities.
3. Photon rings and lensed rings: Additional observable features.
4. Torsion charge: Alters shadow size and intensity.\\
Regarding the (PFDM) model, assuming a perfect fluid:
1. Background geometry: Flat rotation curves hint at dark matter existence and background universe geometry.
2. Energy density and pressure: Expressions for dark matter energy density and pressure can be derived\cite{13,14,15,16,17}. Perfect fluid dark matter is characterized by its uniform density and pressure, which can significantly affect the spacetime around black holes. This influence can lead to observable effects that might alter the expected signatures of the WGC and WCCC. The presence of perfect fluid dark matter could cause the black hole shadow to appear larger or more distorted than predicted in its absence\cite{5000,5001,5002,5003,5004,5005}. This is due to the additional gravitational lensing effect that dark matter exerts on the light passing near the black hole. Dark matter could also impact the emission spectra from the accretion disk of a black hole. The interaction between dark matter particles and ordinary matter could lead to unique spectral lines or shifts in the energy levels of emitted photons\cite{5000,5001,5002,5003,5004,5005}. The merger of black holes embedded in a dark matter halo might produce variations in the amplitude and frequency of the emitted gravitational waves. These variations would be a result of the additional gravitational potential contributed by the dark matter. By analyzing these observational signatures, researchers can compare them with theoretical models that include the effects of perfect fluid dark matter\cite{5000,5001,5002,5003,5004,5005}. Indeed, as mentioned, we initially selected a model that assessed the impact of each observable signature, ensuring that the existence of a black hole structure remains unchallenged. Following this verification, we only investigated the Weak Gravity Conjecture (WGC) to identify a range that preserves all possible conditions while also supporting the establishment of the WCCC. It is crucial to note that the removal of the WGC would result in only a minor alteration in the range related to WCCC compatibility. This implies that, upon validating all the aforementioned cases, a new compatibility—called to as the WGC-can be integrated into this framework. In essence, the WGC may serve as a method to verify the existence and endorsement of the visible signatures we have discussed. In summary, while maintaining all conditions, we endeavored to incorporate a permissible and common interval of WGC into the accptsable range with the WCCC, thereby establishing important compatibility with preserving all structures, as elaborated in the text. These concepts deepen our understanding of the fundamental nature of the universe. Future work could extend this analysis to other types of black holes and dark matter models, as well as explore the possible observational signatures of the Mild WGC and the WCCC.
\begin{center}
\begin{table}
  \centering
\begin{tabular}{|p{4cm}|p{4cm}||p{4cm}||p{4cm}|}
  \hline
  \hspace{1cm} $\gamma < \gamma_{ext}$  & \hspace{1cm} $\gamma = \gamma_{ext}$  & \hspace{1cm} $\gamma > \gamma_{ext}$ & \hspace{1cm} Kind\\[3mm]
   \hline
  \hspace{1cm} No & \hspace{1cm} No & \hspace{1cm} No & \hspace{1cm} $\gamma>0$ \\[3mm]
   \hline
  \hspace{1cm} Yes & \hspace{1cm} Yes & \hspace{1cm} No & \hspace{1cm} $\gamma<0$ \\[3mm]
  \hline
\end{tabular}
\caption{Summary of the results. Yes, indicates the consistency between Mild WGC and WCCC, and No is vice versa.}\label{1}
\end{table}
 \end{center}

\section{Acknowledgments}
The authors would like to express their gratitude to Mohammad Reza Alipour for his insightful discussions. This research is supported by the research grant of the University of Mazandaran (number: 33/26378)

\end{document}